\documentclass[apj]{emulateapj}
\usepackage{mathptmx}

\def\gtorder{\mathrel{\raise.3ex\hbox{$>$}\mkern-14mu
             \lower0.6ex\hbox{$\sim$}}}
\def\ltorder{\mathrel{\raise.3ex\hbox{$<$}\mkern-14mu
             \lower0.6ex\hbox{$\sim$}}}

\slugcomment{Draft of \today}

\shorttitle{GRB~060505}

\shortauthors{Ofek et al.}

\begin{document}

\title{GRB~060505: A possible short-duration gamma-ray burst in a star forming region at redshift of $0.09$}
\author{
E.~O.~Ofek\altaffilmark{1},
S.~B.~Cenko\altaffilmark{1},
A.~Gal-Yam\altaffilmark{1}$^{,}$\altaffilmark{14},
D.~B.~Fox\altaffilmark{3},
E.~Nakar\altaffilmark{1},
A.~Rau\altaffilmark{1},
D.~A.~Frail\altaffilmark{2},
S.~R.~Kulkarni\altaffilmark{1},
P.~A.~Price\altaffilmark{5},
B.~P.~Schmidt\altaffilmark{4},
A.~M.~Soderberg\altaffilmark{1},
B.~Peterson\altaffilmark{4},
E.~Berger\altaffilmark{7}$^{,}$\altaffilmark{8}$^{,}$\altaffilmark{14},
K.~Sharon\altaffilmark{6},
O.~Shemmer\altaffilmark{3},
B.~E.~Penprase\altaffilmark{13},
R.~A.~Chevalier\altaffilmark{11},
P.~J.~Brown\altaffilmark{3},
D.~N.~Burrows\altaffilmark{3},
N.~Gehrels\altaffilmark{9},
F.~Harrison\altaffilmark{1},
S.~T.~Holland\altaffilmark{9},
V.~Mangano\altaffilmark{12},
P.~J.~McCarthy\altaffilmark{7},
D.-S.~Moon\altaffilmark{1},
J.~A.~Nousek\altaffilmark{3},
S.~E.~Persson\altaffilmark{7},
T. Piran\altaffilmark{10},
and R.~Sari\altaffilmark{1}
}
\altaffiltext{1}{Division of Physics, Mathematics and Astronomy, California Institute of Technology, Pasadena, CA 91125, USA}
\altaffiltext{2}{National Radio Astronomy Observatory, PO Box 0, Socorro, NM 87801, USA}
\altaffiltext{3}{Department of Astronomy and Astrophysics, 525 Davey Laboratory, Pennsylvania State University, University Park, PA 16802, USA}
\altaffiltext{4}{Research School of Astronomy and Astrophysics, Mount Stromlo Observatory, via Cotter Road, Weston, ACT 2611, Australia}
\altaffiltext{5}{University of Hawaii, Institute of Astronomy, 2680 Woodlawn Drive, Honolulu, HI 96822-1897}
\altaffiltext{6}{School of Physics and Astronomy and the Wise Observatory, Tel-Aviv University, Tel-Aviv 69978, Israel}
\altaffiltext{7}{Observatories of the Carnegie Institution of Washington, 813 Santa Barbara Street, Pasadena, CA 91101, USA}
\altaffiltext{8}{Princeton University Observatory, Peyton Hall, Ivy Lane, Princeton, NJ 08544, USA}
\altaffiltext{9}{NASA/Goddard Space Flight Center, Greenbelt, MD 20771, USA}
\altaffiltext{10}{Racah Institute of Physics, Hebrew University, Jerusalem 91904, Israel}
\altaffiltext{11}{Department of Astronomy, University of Virginia, PO Box 3818, Charlottesville, VA 22903, USA}
\altaffiltext{12}{INAF, Istituto di Astrofisica Spaziale e Fisica Cosmica di Palermo, Via Ugo La Malfa 153, I-90146, Palermo, Italy}
\altaffiltext{13}{Department of Physics and Astronomy, Pomona College, 610 North College Avenue, Claremont, CA 91711}
\altaffiltext{14}{Hubble Fellow}

\begin{abstract}

On 2006 May 5, a four second duration, 
low energy, $\sim10^{49}$~erg,
Gamma-Ray Burst (GRB) was observed,
spatially associated with a $z=0.0894$ galaxy.
Here, we report the discovery 
of the GRB optical afterglow and observations
of its environment
using Gemini-south, Hubble Space Telescope (HST),
Chandra, Swift and the Very Large Array.
The optical afterglow of this GRB is spatially associated
with a prominent star forming region in
the Sc-type galaxy 2dFGRS~S173Z112.
Its proximity to a star forming region
suggests that the progenitor
delay time, from birth to explosion, is smaller than about 10~Myr.
Our HST deep imaging rules out the presence of a supernova
brighter than an absolute magnitude of about $-11$
(or $-12.6$ in case of ``maximal'' extinction)
at about two weeks after the burst,
and limits the ejected mass of radioactive Nickel 56 to be less
than about $2\times10^{-4}$~M$_{\odot}$ (assuming no extinction).
Although it was suggested that GRB~060505 may belong
to a new class of long-duration GRBs with no supernova,
we argue that the
simplest interpretation is that 
the physical mechanism responsible for this burst is
the same as for short-duration GRBs.

\end{abstract}


\section{Introduction}
\label{Introduction}

Observations of short Gamma-Ray Burst (GRB)
afterglows (e.g. Gehrels et al. 2005;
Fox et al. 2005; Hjorth et al. 2005;
Berger et al. 2005; Bloom et al. 2006)
resulted in a possible dichotomy between
short GRBs and long GRBs -- the presence of a supernova component
in long GRBs and its absence from short GRBs.
However, the recent discovery of GRB~060614, a 102~s-long GRB
with no apparent associated
supernova (Gal-Yam et al. 2006; Fynbo et al. 2006b; Della Valle et al. 2006)
may suggest a more complex picture.
The situation is further tangled as the 
duration distribution of short and long GRBs overlap (e.g. Horv\'{a}th 2002),
and there is no clear way to classify an intermediate
duration GRB based on its duration alone
(despite the fashionable 2~s cut).
Therefore, intensive multi-wavelength observations of nearby
GRBs are needed in order to
construct a clear picture of the GRB zoo,
and to unveil the physical mechanisms behind the
different families of GRBs.

On UTC 2006 May 5, 06:36:01, the Swift Burst Alert Telescope (BAT)
detected the weak GRB~060505 with a fluence of 
$(6.2\pm1.1)\times10^{-7}$~erg~cm$^{-2}$
in the $15-150$~keV band (Palmer et al. 2006; Hullinger et al. 2006),
and a $T_{90}$ duration of $4\pm1$~s.
The gamma-ray time-averaged spectrum was reported to be well fitted by a simple
power law, with an index of $1.3\pm0.3$
(i.e. $dN/dE\propto E^{-1.3}$).
The on-board detection significance for the burst was below the
threshold for an autonomous spacecraft maneuver.
Analysis of the full data set on the ground showed the
burst to be statistically significant
and a repointing was commanded at $\sim0.6$ days.
The Swift X-Ray Telescope (XRT) detected an X-ray
source (Conciatore et al. 2006a)
at a position: $22^{h}07^{m}03.^{s}2$ $-27^{\circ}48'57''$ (J2000.0).
The X-ray position, which has a $90\%$ confidence radius of $4\farcs7$,
is located about $4''$ from the $z=0.0894$ galaxy 2dFGRS~S173Z112
(Colless et al. 2003).
Further observations by Swift XRT,
about five days after the GRB, showed that
the X-ray source decayed between the two epochs
(Conciatore et al. 2006b).
In the optical regime,
Brown \& Palmer (2006) 
did not detect an Optical Transient (OT)
using Swift Ultra-Violet/Optical Telescope (UVOT) observations
of this field conducted $\sim0.6$~d after the GRB trigger.
However,
Ofek et al. (2006) reported the detection
of the OT associated with
GRB~060505 (see \S\ref{Obs}), later confirmed
by VLT/FORS2 observations (Thoene et al. 2006).

In this paper we present multi-wavelength observations
of the afterglow and environment of GRB~060505.
We present our observations in \S\ref{Obs}, 
derive our basic results in \S\ref{Results},
and discuss their implications in \S\ref{Disc}.

\section{Observations}
\label{Obs}

\begin{figure*}
\centerline{\includegraphics[width=16cm]{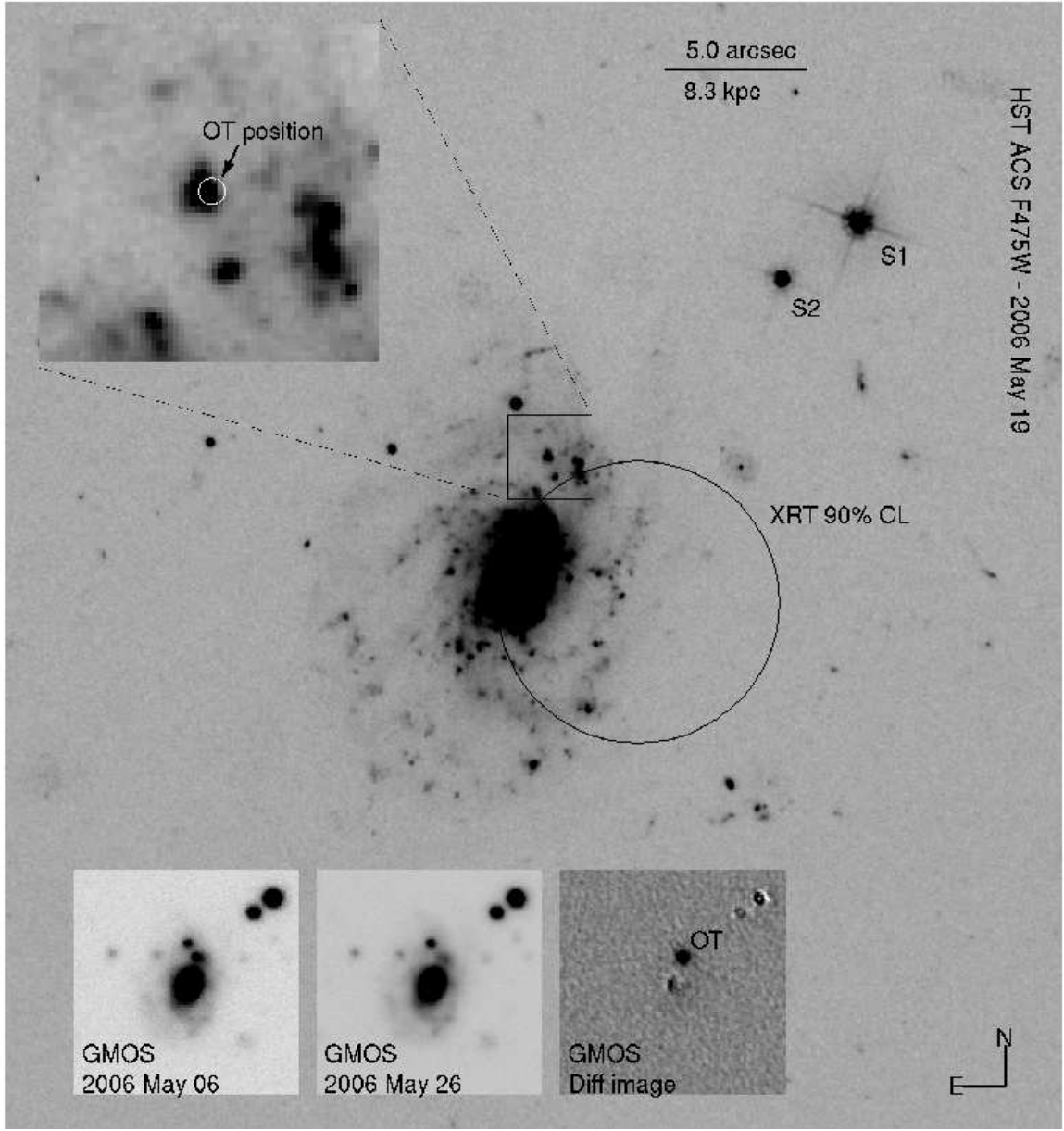}}
\caption{The first epoch of the HST ACS F475W-band image
of the GRB host galaxy 2dFGRS~S173Z112.
The big black circle marks the XRT $90\%$ confidence level (CL)
circle.
S1 and S2 are the reference stars listed in Table~\ref{Tab:Stan}.
The inset at the upper-left corner
zooms in on the position of the OT, which is marked by a white circle.
The radius of the white circle is $0\farcs12$,
which is the 2-$\sigma$ error on the position
of the OT due to the transformation
between the GMOS image and the HST image.
Note that the Chandra X-ray position is consistent with the
OT position.
The insets at the bottom show the g-band GMOS
first epoch image (left), last epoch (middle; see Table~\ref{optical}),
and difference image in which the OT is clearly detected (right).
The OT absolute position,
as measured relative to the USNO-A2 catalog (Monet et al. 1998), is
$22^{h}07^{m}03.^{s}44$ $-27^{\circ}48^{'}51\farcs9$ (J2000.0),
with rms of $0\farcs2$ in each axis.
HST/ACS data were reduced in the standard manner with
IRAF/multidrizzle (Fruchter \& Hook~2002).
\label{HST_Image}}
\end{figure*}

Starting $1.09$~d after the BAT trigger,
we observed GRB~060505 with the Gemini Multi-Object Spectrograph
(GMOS) on the Gemini south telescope
and obtained imaging data at four epochs.
Image subtraction (Alard \& Lupton 1998)
of the first epoch Gemini image from later epochs
have revealed the presence of
an OT (Ofek et al. 2006), spatially associated
with the XRT position, within the
galaxy 2dFGRS S173Z112 at $z=0.0894$
(Colless et al. 2003; see Fig.~\ref{HST_Image}).
The position of the OT is
$22^{h}07^{m}03.^{s}44$ $-27^{\circ}48^{'}51\farcs9$ (J2000.0; rms $0\farcs2$).

Given the possible low-redshift origin of this burst,
we have activated our Hubble Space Telescope (HST)
target of opportunity program, and observed GRB~060505
at two epochs,
14.36 and 32.8 days after the burst,
using the Advance Camera for Surveys (ACS).
In each epoch we integrated for six
orbits with the F475W (SDSS $g$) band, and for three orbits
with the F814W (Cousins $I$) band.
The HST and Gemini images of the OT and host galaxy are presented
in Fig.~\ref{HST_Image}.
The optical afterglow is not detected in the difference images of
the two HST epochs, down to a limiting AB magnitude
of 27.3 and 27.1 in
the F814W and F475W bands, respectively.
It is also not detected by
subtraction of Swift/UVOT images, obtained
at several epochs.

The log of optical imaging observations along with the measurements
and upper limits
are listed in Table~\ref{optical}.
All the limiting magnitudes were obtained by artificially
adding point sources with decreasing magnitudes
to the first epoch images.
To mimic the signal-to-noise properties of the OT,
we placed the artificial stars on top of a star forming region
in the host galaxy, which has a
surface brightness similar to the one in the region where the
OT was found.
Then, we subtracted the latest epoch from each simulated image,
and inspected the images for the artificial sources.
Photometric calibration of the Gemini images 
was based on the HST imaging and
it is presented, with details, in Table~\ref{Tab:Stan}.
%
\begin{deluxetable}{lccccc}
\tablecolumns{6}
\tablewidth{0pt}
\tablecaption{Log of optical observations}
\tablehead{
\colhead{Instrument} &
\colhead{Date} &
\colhead{Exp. Time} &
\colhead{Band} &
\colhead{AB Mag} &
\colhead{Flux} \\
\colhead{} &
\colhead{UTC 2006} &
\colhead{[s]} &
\colhead{} &
\colhead{[mag]} &
\colhead{[$\mu$Jy]}
}
\startdata
\hline
GMOS   & 05-06.377 & $5\times180$ & r        & $21.93\pm0.16$ & $6.1\pm1.0$  \\
GMOS   & 05-06.393 & $5\times240$ & g        & $22.43\pm0.08$ & $3.9\pm0.3$  \\
GMOS   & 05-12.316 & $5\times180$ & g        & $>24.8$        & $<0.43$      \\
GMOS   & 05-12.330 & $5\times180$ & r        & $>24.3$        & $<0.69$      \\
GMOS   & 05-14.353 & $5\times180$ & g        & $>24.6$        & $<0.52$      \\
GMOS   & 05-14.367 & $5\times180$ & r        & $>24.7$        & $<0.48$      \\
GMOS   & 05-14.380 & $5\times180$ & i        & $\gtorder23.3$ & $\ltorder1.7$\\
GMOS   & 05-26.318 &$10\times300$ & g        & \nodata\tablenotemark{b}  & \nodata      \\
GMOS   & 05-26.294 & $1\times300$ & r        & \nodata\tablenotemark{b}  & \nodata      \\
\hline
UVOT\tablenotemark{a}& 05-05.981 & $834.7$& U& $>21.3$        & $<10.7$       \\
UVOT   & 05-10.512 &  $1300.9$    & U        & \nodata\tablenotemark{b}  & \nodata      \\
UVOT   & 05-05.996 &   $834.6$    & V        & $>20.4$        & $<53$        \\
UVOT   & 05-21.818 & $11476.5$    & V        & \nodata\tablenotemark{b}  & \nodata      \\
UVOT   & 05-05.977 &  $1666.3$    & UW1      & $>22.1$        & $<5.8$       \\
UVOT   & 05-18.241 &  $1735.6$    & UW1      & \nodata\tablenotemark{b}  & \nodata      \\
UVOT   & 05-05.999 &  $2228.9$    & UM2      & $>23.7$        & $<3.1$       \\
UVOT   & 05-10.505 &  $3974.5$    & UM2      & \nodata\tablenotemark{b}  & \nodata      \\
\hline
ACS    & 05-19.635 & $9\times783$ & F814W    & $>27.3$        & $<0.044$     \\
ACS    & 05-19.636 &$18\times783$ & F475W    & $>27.1$        & $<0.053$     \\
ACS    & 06-06.957 & $9\times760$ & F814W    & \nodata\tablenotemark{b}  & \nodata      \\
ACS    & 06-07.160 &$18\times760$ & F475W    & \nodata\tablenotemark{b}  & \nodata      \\
\enddata
%
\tablenotetext{a}{Swift UVOT Vega-based zero points used are
$17.29$, $17.69$, $18.38$, and $17.88$ for the
UM2, UW1, U, and V-bands, respectively. To convert Vega-based magnitudes to AB magnitudes
we added $1.65$, $1.39$, $0.99$, and $0.00$ magnitudes to the $UM2$, $UW1$, $U$, and $V$-bands, respectively.}
\tablenotetext{b}{Used as a reference image in the image subtraction process.}
\label{optical}
\end{deluxetable}
\begin{deluxetable}{lcccccc}
\tablecolumns{6}
\tablewidth{0pt}
\tablecaption{Reference Stars}
\tablehead{
\colhead{Name} &
\colhead{R.A.} &
\colhead{Dec.} &
\colhead{g} &
\colhead{r} &
\colhead{i} \\
\colhead{} &
\multicolumn{2}{c}{J2000.0} &
\multicolumn{3}{c}{AB magnitude\tablenotemark{a}}
}
\startdata
S1\tablenotemark{b} & 22:07:02.95 & $-$27:48:44.4  & $19.70\pm0.02$ & $19.35\pm0.02$ & $19.26\pm0.02$ \\
S2\tablenotemark{c} & 22:07:03.15 & $-$27:48:46.5  & $20.95\pm0.03$ & $19.61\pm0.06$ & $19.16\pm0.02$ \\
\enddata
\tablenotetext{a}{The magnitudes were calculated by fitting 
the synthetic photometry magnitudes of
stellar spectral templates (Pickles 1998) to the observed ACS F475W and F814W-band magnitudes, and then calculating the synthetic magnitudes in the $g$, $r$, and $i$-bands using the best fit spectral template. The ACS magnitudes were measured in a four-pixel radius aperture and were extrapolated to an infinite aperture (Sirianni et al. 2005).}
\tablenotetext{b}{The best fit spectral template for this object is of an F8V star. The ACS F475W magnitude is $19.69\pm0.02$, and the ACS F814W magnitude is $19.24\pm0.02$.}
\tablenotetext{c}{The best fit spectral template for this object is of an K3I star. The ACS F475W magnitude is $20.90\pm0.03$ and the ACS F814W magnitude is $19.04\pm0.02$.}
\label{Tab:Stan}
\end{deluxetable}
%
%
%

\begin{figure}
\centerline{\includegraphics[width=8.5cm]{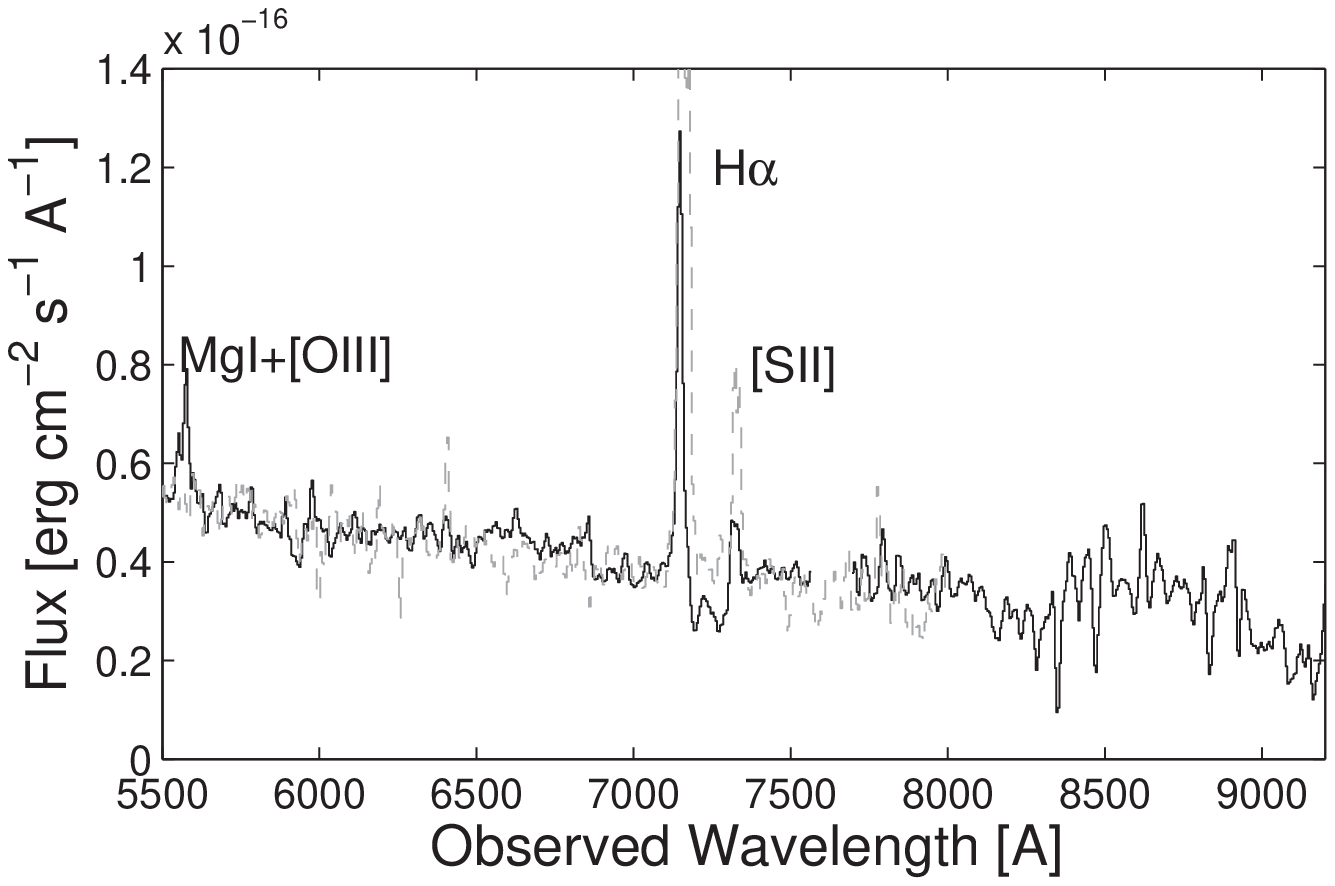}}
\caption{The spectrum of the host galaxy of GRB~060505
obtained by the DBSP mounted on the Palomar observatory 200-inch
telescope (solid black line).
The spectrum continuum shape and emission lines,
as well as the galaxy morphology (Fig.~\ref{HST_Image})
are typical for an Sc galaxy.
However, we note that we do not identify
the absorption features redward of the H$\alpha$ line.
A scaled spectrum template of an Sc galaxy (Kinney et al. 1996)
redshifted to match the host galaxy
is shown as the gray dashed line.
The gap around 7600~\AA~is due to removal of a Telluric absorption.
\label{HostGalSpec_P200red}}
\end{figure}

On 2006 May 13,
we obtained a $2\times30$~min spectrum of the position
of the OT using the Gemini-south
telescope with the GMOS instrument.
We used a $1''$ slit with the R400 grating
blazed at 6000~\AA.
The spectrum shows a feeble continuum emission
and prominent emission lines at the redshift
of the host galaxy, confirming
that the bright knot at the OT position is indeed
a star forming region within the galaxy
(see also Fynbo et al. 2006b).
In order to estimate the star formation rate
within the host galaxy, on 2006 July 19 we obtained a
flux calibrated spectrum of
the central part of the host galaxy using the
Double Beam Spectrograph (DBSP) on the Palomar 200-inch telescope.
We used a $1''$ slit with the R158 grating blazed at 7500~\AA.
The red-arm spectrum, shown in Fig.~\ref{HostGalSpec_P200red},
consists of two 15~min exposures.
An H${\alpha}$ emission line is clearly detected
in the spectrum, indicating ongoing star formation in
this galaxy.
%
%

At about $19.2$~d after the burst,
we observed GRB~060505 using the ACIS-S detector on board
the Chandra X-ray observatory.
Using wavedetect\footnote{part of the Chandra Interactive Analysis of Observations software.},
we identified a source consistent with the OT position.
The source was detected at the $3.5$-$\sigma$ confidence level
in the $2$--$8$~keV band.
The XRT and Chandra
X-ray measurements are listed in Table~\ref{Table-Xray}.
%
\begin{deluxetable}{ccccccc}
\tablecolumns{7}
\tablewidth{0pt}
\tablecaption{Log of X-ray observations}
\tablehead{
\colhead{Date\tablenotemark{a}} & 
\colhead{Exp.} &
\colhead{Band\tablenotemark{c}} &
\colhead{Count rate} &
\colhead{$\Gamma$\tablenotemark{d}} &
\colhead{Flux\tablenotemark{e}} &   
\colhead{Flux\tablenotemark{f}} \\   
\colhead{Tel.\tablenotemark{b}} &
\colhead{[ks]} &
\colhead{[keV]} &
\colhead{[count~s$^{-1}$]} &
\colhead{} &
\colhead{[erg~s$^{-1}$~cm$^{-2}$]} &
\colhead{[nJy]} 
}
\startdata
05.973 & 8.0 &0.2--10 & $(1.1\pm0.1)$         & 2.0 & $(4.6\pm0.4)\times10^{-13}$ & $49.2$ \\
Swift  &     &        & $\times10^{-2}$       & 2.5 & $(4.2\pm0.4)\times10^{-13}$ & $45.5$ \\
\hline
10.031 &11.7 &0.2--10& $(8.6\pm4.0)$         & 2.0 & $(3.6\pm1.7)\times10^{-14}$ & $3.84$ \\
Swift  &     &        & $\times10^{-4}$       & 2.5 & $(3.3\pm1.5)\times10^{-14}$ & $3.55$ \\
\hline
24.334 &24.71&2.0--8  & $(1.2_{-0.9}^{+1.6})$ & 2.0 & $(5.4_{-3.9}^{+6.9})\times10^{-15}$ &$0.571$\\
Chandra&     &        & $\times10^{-4}$       & 2.5 & $(9.4_{-6.7}^{+12})\times10^{-15}$  &$0.997$ \\
%
%
\enddata
\tablenotetext{a}{UTC 2006 May, time of beginning of observation.}
\tablenotetext{b}{The instrument used is listed below the date.}
\tablenotetext{c}{Band for the X-ray counts.}
\tablenotetext{d}{Photon index assumed in the conversion from counts to flux.}
\tablenotetext{e}{Unabsorbed flux in the 0.2--10~keV band. Assuming Galactic neutral Hydrogen column density of $N_{H}=1.8\times10^{20}$~cm$^{-2}$ (Dickey \& Lockman 1990).}
\tablenotetext{f}{Specific flux at 1keV.}
\tablecomments{Swift-XRT X-ray counts are from Conciatore et al. (2006b).}
\label{Table-Xray}
\end{deluxetable}
%
The X-ray flux was calculated using
WebPIMMS\footnote{http://cxc.harvard.edu/toolkit/pimms.jsp},
assuming the X-ray spectrum is described by a power-law
with a photon index of $2.0$, and a Galactic
Hydrogen column density of $N_{H}=1.8\times10^{20}$~cm$^{-2}$
(Dickey \& Lockman 1990).

\begin{figure}
\centerline{\includegraphics[width=8.5cm]{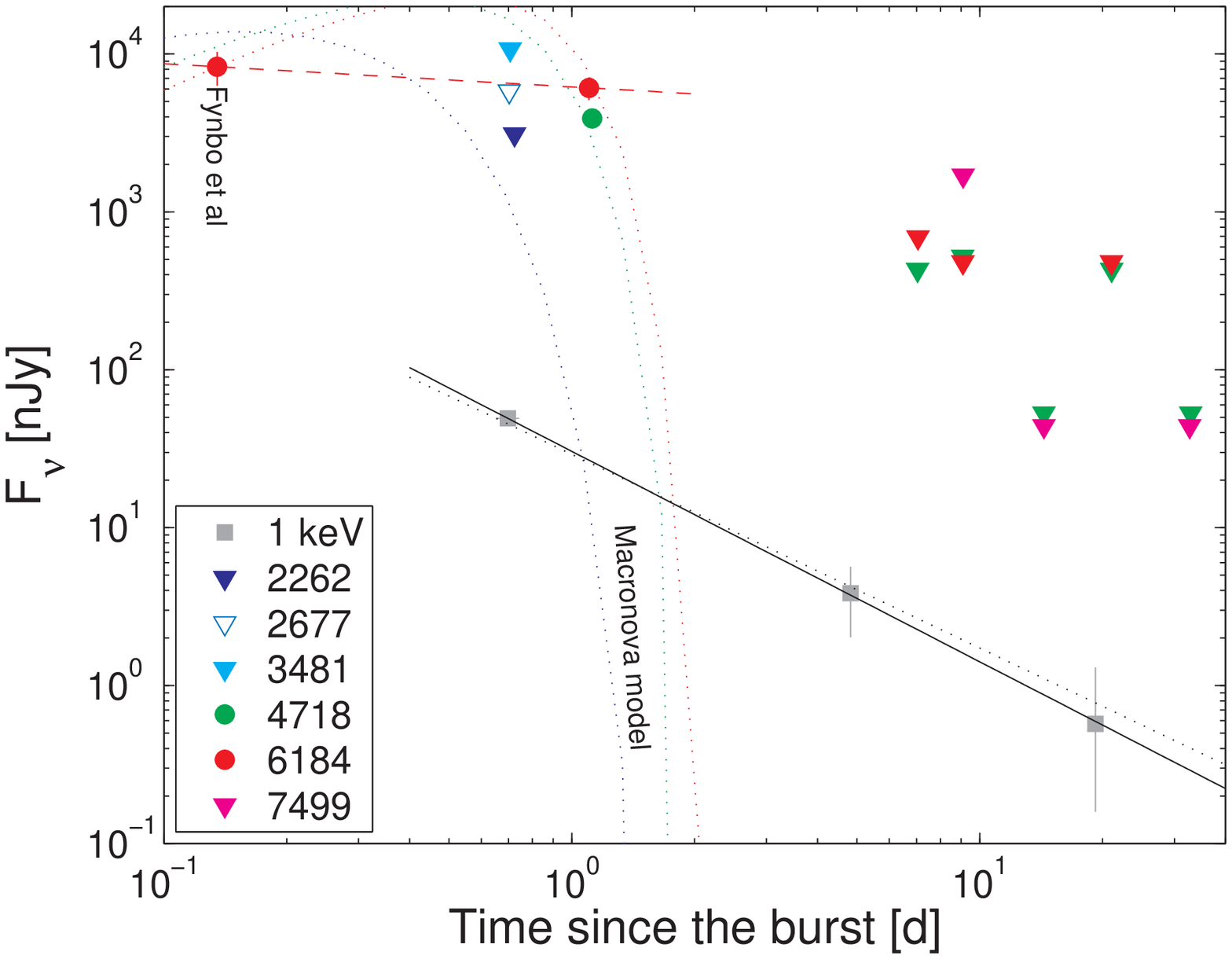}}
\caption{X-ray (gray squares) and optical (color circles and triangles)
light curves of GRB~060505
(see Tables~\ref{optical} and \ref{Table-Xray}).
The circles represent the optical measurements, while the triangles
mark upper limits.
The wavelength of each measurement is color coded and shown
in the legend (in Angstroms or keV),
where $2262$, $2677$, $3481$, $4718$, $6184$ and $7499$,
corresponds to $UM2$, $UW1$, $U$, $g$, $r$, and $i$-bands, respectively.
The dashed-red line shows a power-law with decay index of $-0.15$
that fits the two early $r$-band measurements.
The red, green and blue dotted lines show
a macronova model
with an ejecta mass of $3\times10^{-2}\,$M$_{\odot}$ and
ejecta velocity of $0.6$ the speed of light,
in the $UM2$ (2262~\AA), $g$ (4718~\AA), and $r$ (6184~\AA)-bands,
respectively (see Kulkarni 2005 for details).
In the conversion of the X-ray counts to flux we assumed
the X-ray spectrum is described by a power-law with
a photon index of $2.0$.
The solid-black line shows the best-fit power-law
to the X-ray data, with an
index $\alpha_{X}=-1.33\pm0.17$ ($\chi^{2}/dof=0.005/1$).
Assuming a power-law with a photon index of $2.5$
gives $\alpha_{X}=-1.22\pm0.16$ ($\chi^{2}/dof=0.3/1$; dotted-black line).
\label{GRB060505_XrayOptLC}}
\end{figure}

Figure~\ref{GRB060505_XrayOptLC} shows the X-ray
and optical light curves of GRB~060505.
The X-ray light curve is well fitted by
a single power-law with a slope of $\alpha_{X}=-1.33\pm0.17$ (solid line).
Our early $r$-band measurement together with
the single $R$-band data presented in
Fynbo et al. (2006b; converted to the AB magnitude system, $r_{AB}=R_{Vega}+0.23$),
suggest a power-law decay rate of about $-0.15\pm0.14$ (dashed line).
%
%

We have observed GRB~060505 with the 
Very Large Array\footnote{The Very Large Array is operated
by the National Radio Astronomy Observatory, a facility of the National
Science Foundation operated under cooperative agreement by
Associated Universities, Inc.},
at frequency of $8.46$~GHz and 100~MHz bandwidth,
at the following epochs:
UTC 2006 May 17.53; Aug 18.32, 20.35, 21.32, 22.31, and 23.37.
The reduction was done 
following standard practice in the
Astronomical Image Processing System software package.
At the first epoch, 12.3~d after the burst,
no radio source was visible at
the OT position to a 3-$\sigma$ limit of 165 $\mu$Jy.
Summing the last five epochs,
with mean epoch 107.8~d after the burst,
reveal no radio source
to a 3-$\sigma$ limit of 46 $\mu$Jy at the OT position.

\section{Results}
\label{Results}

%
At $z=0.0894$, the luminosity
distance\footnote{Assuming WMAP$+$SDSS cosmological parameters: $H_{0}=70.9$~km~s$^{-1}$~Mpc$^{-1}$; $\Omega_{m}=0.266$; $\Omega_{\Lambda}=1-\Omega_{m}$ (Spergel et al. 2006).}
to GRB~060505 is 404~Mpc
(distance modulus $38.03$~mag;
angular diameter distance 341~Mpc).
The relatively low redshift of GRB~060505
makes it an extraordinary event for both constraining
the presence of a supernova and the ejecta from the explosion,
and for studying the GRB environment.
We discuss these issues below.

\subsection{Environment and progenitor age}
%
%
As seen in Fig.~\ref{HST_Image}, GRB~060505
is spatially associated with a bright,
spectroscopically confirmed,
star forming region found
at a projected distance of $7.1$~kpc ($4\farcs3$)
from its host galaxy center.
The AB magnitude of this star forming knot is
23.9 and 23.7 in the F475W and F814W-bands, respectively.

Our observations
(Fig.~\ref{HST_Image} and \ref{HostGalSpec_P200red})
show that 2dFGRS~S173Z112 is a face-on Sc-type galaxy.
Using the flux of the H$\alpha$ emission line,
along with the relation from Kennicutt (1998),
we estimate that the star formation rate in this galaxy
is about 2~M$_{\odot}$~yr$^{-1}$.
We obtained this value by extrapolating
the H$\alpha$ emission, observed within the slit,
over the entire area of the galaxy, weighted by the g-band surface brightness.
%
%
We note that the total AB magnitude of the galaxy
is 18.3 and 17.7 in the F475W and F814W-bands, respectively.
This is equivalent to absolute magnitude of $-19.4$ and $-20.4$
in the F475W and F814W-bands, respectively (corrected for K-correction; Oke \& Sandage 1968).
%
%
%
Therefore, the specific star formation in this galaxy
is $\sim3$~M$_{\odot}$~yr$^{-1}$~L$_{*}^{-1}$.

The association of GRB~060505 with a bright star forming region
can be used to set a limit on the delay time,
from progenitor birth to explosion.
This limit may be especially interesting if GRB~060505
is a consequence of a compact-binary merger rather than
due to a massive-star core collapse.
First, the presence of an HII
region in the explosion site of GRB~060505 suggests
a delay time below about 10~Myr (e.g. Mayya 1995).
Second, a limit on the delay time that is applicable for
the compact-star merger scenario can be derived
from the size of the star forming region combined
with the speed of the progenitor.
The diameter of
$d\sim400$~pc of the HII region at the position of the OT
suggests that the delay time, $\tau$,
from birth to explosion of GRB~060505 is:
\begin{equation}
\tau \ltorder 11\times10^{6} \frac{d}{400 {\rm pc}} \left( \frac{v}{35 {\rm km~s^{-1}}} \right) ^{-1} {\rm yr},
\label{Eq-Tau}
\end{equation}
where $v$ is the (kick) velocity of the progenitor.
The $35$~km~s$^{-1}$ kick velocity we adopted
is half 
(in order to account for the binary center of mass speed)
of the value of the lowest transverse velocity
of a pulsar in the sample of Hansen \& Phinney (1997).

\subsection{Supernova Limits}
%
%
The absence of detectable optical emission
from the OT at $\gtorder2$~d
after the burst (see Table~\ref{optical}),
suggests that GRB~060505 was
not associated with a bright supernova.
Assuming 
negligible extinction\footnote{The Galactic extinction toward GRB~060505 is $E_{B-V}=0.02$ (Schlegel et al. 1998).},
our deep HST observations imply 
a limit on the $g$ and $i$-bands absolute magnitude of a supernova,
two weeks after the burst, of $\gtorder-11$ mag.
Extinction within the host galaxy
may weaken our results.
However, the Balmer lines ratio of the star formation region
at the position of the OT is suggestive
of no reddening (Fynbo et al. 2006b).
We note that the color
of the afterglow ($f_{\nu}\propto \nu^{-1.7\pm0.7}$),
is somewhat redder than expected for GRB afterglow.
Therefore, it may suggest some extinction.
We put a conservative upper limit on the extinction
by using the observed color of the OT at one day
after the burst and by assuming
that the intrinsic optical spectrum
cannot be steeper than a Rayleigh-Jeans spectrum ($f_{\nu}\propto \nu^{2}$).
We find that $E_{B-V}<1.04$
(assuming $A_{V}/E_{B-V}=3.08$; Cardelli et al. 1989).

Assuming this ``maximal extinction'',
the $i$-band absolute AB magnitude of a supernova
lurking in the OT position
is fainter than $-12.6$.
This limit includes a K-correction
(Humason, Mayall, \& Sandage 1956; $m_{int}=m_{obs}-K$)
of $0.05$~mag in the $i$-band,
assuming an SN1998bw-like spectrum.
Our limit improves upon the result of Fynbo et al. (2006b)
by three magnitudes (at about two weeks after the burst),
and rules out even the faint class of supernova they discuss.

%
Furthermore, using our HST limit on day 14, along with
Eq.~44 in Kulkarni (2005)
we can place an approximate upper limit on the 
mass of the radioactive Nickel~56 produced in this explosion
of M$_{^{56}Ni}\ltorder2\times10^{-4}$~M$_{\odot}$,
assuming no extinction and
M$_{^{56}Ni}\ltorder10^{-3}$~M$_{\odot}$, assuming maximal extinction.
We note that the faintest core collapse supernovae known to date,
ejected about $(2-8)\times10^{-3}$~M$_{\odot}$ of $^{56}$Ni
(e.g. Pastorello et al. 2004), but there may be a bias
against finding such low-luminosity SN.

\subsection{The Light Curve and Spectral Energy Distribution}

As shown in Fig.~\ref{GRB060505_XrayOptLC},
during the first day, the optical light curve evolution
is consistent with a power-law decay
with a $-0.15\pm0.14$ index.
The photometry was performed after image subtraction,
so this result is not sensitive to
contamination by the host galaxy light.
Such a flat temporal evolution
is rare among optical GRB afterglows.
Interestingly, similar
flat temporal optical evolution
was observed in the short-duration GRB~060313
(Roming et al. 2006).
However, we note that this power-law decay is based on
sparsely sampled light curve.
At later times, between day one and 14,
the power-law index is steeper than about $-1.9$.
The X-ray light curve, on the other hand, is
consistent with a single
power-law decay rate of $\alpha_{X}=-1.33\pm0.17$.

Although the optical-to-X-ray spectral power-law index
$\beta_{ox}=-0.80\pm0.03$
(defined by $f_{\nu}\propto \nu^{\beta_{ox}}$)
that we measured at 1.1 days after the burst 
is typical for GRB afterglows,
the visible-light color of the afterglow
($f_{\nu}\propto\nu^{-1.7\pm0.7}$)
is marginally redder than
expected at one~day after the burst
(e.g.~{\v S}imon et al. 2001; Lipkin et al.~2004).
Possible explanations are
either that the OT is reddened
or that some of the optical radiation is contributed
by some additional mechanism (i.e. not by the afterglow).
An interesting possibility is that
the early optical emission is powered,
in addition to the afterglow light,
by the decay of free neutron ejecta.
Such a scenario was suggested
in the context of neutron star (NS) mergers
by Li \& Paczy{\'n}ski (1998), and investigated
by Kulkarni (2005; i.e. macronova).
In Fig.~\ref{GRB060505_XrayOptLC}
we show, for example, a macronova model that roughly fits
the optical data. The dotted red, green, and blue
lines correspond to a macronova model
with $3\times10^{-2}$~M$_{\odot}$ free-neutron ejecta
and a velocity of $0.6$ the speed of light.
%
An inspection of the models presented by Kulkarni (2005)
suggests that the radioactive decay of the
free neutron ejecta may roughly explain the
behavior of the optical light curve.
However, our observations are too sparse for a critical
examination of this model.

\section{Discussion}
\label{Disc}

GRBs are traditionally classified according to their durations
into short and long events (Kouveliotou et al. 1993).
GRB~060505 has some unusual characteristics
that make it difficult to place within this scheme.
Fynbo et al. (2006b) suggested that GRB~060505 may belong
to a new emerging group of long duration GRBs
without a supernova.
We note that
the existence of a third group of
GRBs was already suggested in the past
(e.g. Mukherjee et al. 1998; Horv\'{a}th 2002), based on the analysis
of the GRB duration distribution.
Recently, a third group was also discussed
by Gal-Yam et al. (2006),
Fynbo et al. (2006b), and Della Valle et al. (2006)
which presented observations of GRB~060614 --
a long-duration GRB ($T_{90}\approx102$~s; Barthelmy et al. 2006),
spatially associated with a $z=0.125$ galaxy,
with no apparent supernova.
However, a simpler explanation for this particular case
is that GRB~060505 was a short-duration GRB.
We note that, ``short'' is used here in the sense that
the GRB physical mechanism is similar to that of short GRBs.
We discuss the various possibilities
for the nature of GRB~060505 below.


The absence of a supernova, the low
isotropic equivalent $\gamma$-ray energy,
$E_{\gamma,iso}=(1.2\pm0.2)\times10^{49}$~erg,
and the low-redshift of GRB~060505,
are characteristic of other short-duration GRBs
(see however Berger et al. 2006).
On the other hand,
the duration of GRB~060505
is above the canonical 2~s cut,
separating short and long GRBs.
However, the duration distributions of
short and long GRB
overlap (e.g. Horv\'{a}th 2002; Donaghy et al. 2006).
For example,
adopting the Horv\'{a}th (2002) decomposition (based on BATSE observations),
the probability of a short burst
to have a $T_{90}$ duration 
larger than 4~s is about $12\%$.
Moreover, $64\%$ of the
GRBs with $\sim4$~s durations belong
to the ``short''-duration group.
Although the Horv\'{a}th (2002) decomposition is not 
necessarily the correct one, it is preferred over
using a sharp cutoff at 2~s.

If indeed GRB~060505 is a genuine
short-duration
GRB, then our limit on the delay time implies that any scenario
for short-duration GRBs should be able to produce
events with delay times shorter than about 10~Myr.
In the context of the NS merger scenario,
we note that
Belczynski et al. (2006) estimated that
the time delay from binary-birth-to-explosion probability function
has a narrow maximum
around 20~Myr, with approximately 10~Myr width, followed
by a flat probability extending to delay times of the order
of the Hubble time.
Furthermore,
NS-black hole (BH) mergers may have somewhat
shorter delay times than those
of NS--NS mergers.
Therefore, if indeed GRB~060505 was a short-duration GRB,
our observations are consistent with a
NS--NS/NS--BH merger model for short GRBs.


Next we address the possibility that GRB~060505 is a long-duration GRB.
The redshift and the isotropic equivalent $\gamma$-ray energy of GRB~060505
are atypically low for a ``classical'' long GRB.
While several anomalous
low energy, local Universe, long GRBs
have been detected (e.g. GRB060218; Campana et al. 2006),
GRB~060505 and GRB~060614 (Gal-Yam et al. 2006)
are not associated with a supernova.

%
Following 
Fruchter et al. (2006)
we compared the environment of GRB~060505
to that of other long GRBs.
Fruchter et al. (2006)
studied a sample of long GRBs with identified host galaxies.
They
sorted all the pixels in each host galaxy image
by brightness, and computed the fraction
of the total light of the host contained
in pixels fainter than or equal to the pixel
in which the OT is located. Next, they presented
a histogram of this fraction of light,
and found that long GRBs prefer to reside
in the brightest pixels of a galaxy.
We applied the Fruchter et al. (2006) analysis to an
F475W-band image of GRB~060505.
Using exactly the same SExtractor
(Bertin \& Arnouts 1996)
detection criteria used 
by these authors, we find that the brightness
at the spatial position of the OT is
among the dimmest first-to-$45$th percentile
(depending on the exact location of the OT, which is known to
about 60~mas) of the galaxy light.
Comparing this to the long-duration GRBs population investigated
by Fruchter et al. (2006), we find that 
fewer than $10\%$ of Fruchter et al. (2006) long-duration GRBs
occurred in similar (or fainter) regions to that of GRB~060505.
We note that our analysis may be biased
by two facts.
First, our observations were made at rest-frame wavelengths
of about 4400~\AA, while the Fruchter et al. (2006)
analysis was performed at a median
rest-frame wavelength of about 3700~\AA.
However, as shown by Fruchter et al. (2006),
their analysis is not sensitive to the redshift
of the galaxies in their sample.
Second, the redshift of GRB~060505 is smaller (by a factor $\sim9$)
than the typical redshift of the GRBs
in the Fruchter et al. (2006) sample.
Placing our galaxy at the typical redshift of the Fruchter et al. sample
will bring most of the galaxy light below the noise level.
Hence, the fraction
of observable galaxy light
that is dimmer than the brightness at the OT position decreases,
making GRB~060505 less consistent with the long-duration GRBs population.
We note that a possible interpretation of the Fruchter et al.
analysis is that long GRBs are associated with low metallicity
environment (see however, Fynbo et al. 2006a).
Therefore, long GRBs in massive spirals are expected
to occur at the outskirts of the galaxy, where the metallicity is
on average lower.
In this case the fact that GRB~060505
occurred in the outer region of its host galaxy
may support the notion that it is a long GRB (although
until now long GRBs were not observed in massive spirals).
Therefore, it will be constructive to measure the metallicity
of GRB~060505 host galaxy and the immediate environment of the GRB
and to compare it with the metallicity of other long GRB hosts.

%

Fynbo et al. (2006b) suggested that both GRB~060505 and GRB~060614
may constitute a new class of GRBs, which 
are associated with faint supernovae
(e.g.  Fryer, Young, \& Hungerford 2006).
If a class of long GRBs with faint (or no) supernovae
exists, a clue regarding the nature of the progenitor
can be derived from the rate of such events,
which we derive below.
To estimate the rate
we took into account 
the BAT field of view ($1.4$~sr) and operation time ($1.67$~yr),
and assumed that at least two such bursts were observed by
Swift/BAT.
Next, we summed
the volume to the two GRBs and multiply it by two.
We introduced the factor of two 
in order to account for the fact that we can detect
bursts at larger distances.
It has been shown (Schmidt 1968) that
the expectancy value for the
ratio of the volume enclosed within the distance
to an event
to the maximal volume in which such an event is detectable
(i.e. $V/V_{max}$) is $1/2$, hence the factor of 2.
We get that
the local rate of such GRBs is
$\gtorder1.5$~yr$^{-1}$~Gpc$^{-3}$, at the $95\%$ confidence level (Gehrels 1986).
This lower limit is approximately three times larger
than the observed rate of classical long GRBs
(without beaming correction; e.g. Schmidt 2000; Guetta, Piran, \& Waxman 2005).
However, this limit is below the inferred rate of
low-luminosity long GRBs such as GRB~980425 and
GRB~060218 (e.g., Soderberg et al. 2006).

Finally, we discuss the possibility
that GRB~060505 is a background event unrelated to
the galaxy 2dFGRS~S173Z112.
Schaefer \& Xiao (2006) claimed that the association
of GRB~060505 with the galaxy 2dFGRS~S173Z112
is due to a chance coincidence.
Estimating the probability for a chance coincidence
is susceptible to the pitfalls of aposteriori statistics.
Moreover, it is not clear how to incorporate
additional information like the fact that GRB~060505 is
associated with a star forming region.
Bearing this in mind,
we note that in their calculation Schaefer \& Xiao (2006)
used an inappropriate luminosity
(of $0.03$~L$_{*}$) for the host galaxy of GRB~060505.
Repeating their calculation using the correct value ($0.67$~L$_{*}$),
and using the galaxy luminosity function given by Blanton et al. (2003),
we get that the probability for a chance coincidence per trial
is six times lower than the probability of $0.8\%$
claimed by Schaefer \& Xiao (2006).
While a chance projection cannot be definitely ruled out for
this burst, its association with 2dFGRS~S173Z112 is no less certain than
those of prototypical short GRBs (e.g., GRB 050709 and GRB 050724)
with their host galaxies.

To summarize,
our observations show that GRB~060505 is not
associated with any supernova brighter than absolute $i$-band magnitude
of $-11$ (assuming no extinction).
Furthermore, we can constrain the mass of radioactive Nickel 56 ejecta
to be less than $2\times10^{-4}$~M$_{\odot}$, and we find
that the delay time, from birth to explosion,
of this GRB progenitor is probably smaller than about $10$~Myr.
The light curve of GRB~060505 is atypical for GRB afterglows
and we show it may be explained by the macronova model
(Kulkarni 2005).
Alternatively, the OT may be somewhat reddened.
The simplest interpretation of GRB~060505
is that it is the nearest observed short-duration GRB to date.
Another possibility is that
GRB~060505 belongs to a new class
of supernova-lacking GRBs that have long durations.
However, this will require that the rate of
such events will be at least a factor of three
larger than that of ``classical'' long GRBs.

\acknowledgments
EOO thanks Orly Gnat
for valuable discussions
and to an anonymous referee for useful comments.
This work is supported in part by grants from NSF and NASA.

\end{document}